\documentclass[%
    secnumarabic,
    superscriptaddress,
    preprint,
    preprintnumbers,
    amsmath,
    amssymb,
    longbibliography,
    tightlines
    ]{revtex4-1}
\usepackage{mathtools}
\usepackage{bbm}
\usepackage{commath}
\usepackage{libertine}
\usepackage[english]{babel}

\usepackage{graphicx}
\usepackage{dcolumn}
\usepackage{bm}
\usepackage{hyperref}
\usepackage[mathlines]{lineno}

\usepackage{todonotes}
\usepackage{verbatim}
\usepackage{placeins}

\usepackage{derivative}

\begin{document}


\title{An Agent-Based Extension to Sector-Wise Input-Output Recovery Models}

\author{Jan Hurt}
\affiliation{Complexity Science Hub Vienna, Vienna, Austria}

\author{Stefan Thurner}
\affiliation{Section for Science of Complex Systems, CeDAS, Medical University of Vienna, Vienna, Austria}
\affiliation{Complexity Science Hub Vienna, Vienna, Austria}
\affiliation{Santa Fe Institute, Santa Fe, NM, USA}
\affiliation{Supply Chain Intelligence Institute Austria (ASCII), Metternichgasse 8, 1030 Vienna, Austria}

\author{Peter Klimek}
\affiliation{Section for Science of Complex Systems, CeDAS, Medical University of Vienna, Vienna, Austria}
\affiliation{Complexity Science Hub Vienna, Vienna, Austria}
\affiliation{Supply Chain Intelligence Institute Austria (ASCII), Metternichgasse 8, 1030 Vienna, Austria}
\affiliation{Department of Clinical Neuroscience, Division of Insurance
Medicine, Karolinska Institutet, Stockholm, Sweden}

\date{\today}
\begin{abstract}
Dynamic input-output models are standard tools for understanding inter-industry dependencies and how economies respond to shocks like disasters and pandemics. However, traditional approaches often assume fixed prices, limiting their ability to capture realistic economic behavior. Here, we introduce an adaptive extension to dynamic input-output recovery models where producers respond to shocks through simultaneous price and quantity adjustments. Our framework preserves the economic constraints of the Leontief input-output model while converging towards equilibrium configurations based on sector-specific behavioral parameters. When applied to input-output data, the model allows us to compute behavioral metrics indicating whether specific sectors predominantly favor price or quantity adjustments. Using the World Input-Output Database, we identify strong, consistent regional and sector-specific behavioral patterns. These findings provide insights into how different regions employ distinct strategies to manage shocks, thereby influencing economic resilience and recovery dynamics.
\end{abstract}


\maketitle

\section{%
    Introduction
}
Modeling the propagation of and recovery from economic shocks is an extremely relevant endeavor in macroeconomics and policy analysis \cite{acemogluNetworksMacroeconomyEmpirical2016}. Severe shocks, such as those triggered by financial crises, natural disasters, or pandemics, often generate significant disruptions in interconnected industries. Understanding how these shocks spread through production networks is essential for designing resilient and effective policy responses.

Traditional models of economic shock propagation are often based on input–output (IO) economics, where economies are represented as networks where nodes correspond to industrial sectors and edges represent the flow of goods between them (e.g., \cite{acemogluNetworksMacroeconomyEmpirical2016, millerInputOutputAnalysisFoundations2009}).
In these IO models, the following is typically assumed: the IO matrix  describes intersectoral flows, technical coefficients are constant, and monetary inflows match outflows for every sector.

Various studies adapted these core IO assumptions through different adaptive processes to address diverse economic challenges. For example, \citet{hallegatteAdaptiveRegionalInputOutput2008} employs an adaptive regional IO model to assess the impact of natural disasters on regional production by integrating supply-side limitations such as production bottlenecks and underproduction. Similarly, \citet{contrerasPropagationEconomicShocks2014} explore modifications of the classic Leontief model, focusing on quantity adjustments in shock diffusion, while \citet{klimekQuantifyingEconomicResilience2019a} uses a dynamic Leontief model with constant coefficients to estimate recovery processes following financial crises. \citet{pichler21} assess the impacts of COVID-19 lockdowns using a framework based on Leontief production functions to capture production bottlenecks.
The input-output inoperability model (IIM) proposed in \citet{haimesInoperabilityInputOutputModel2005} introduces an “industry resilience coefficient” to capture the ability of sectors to cope with disruptions, thus extending the classic framework by considering the inability of a sector to function as intended. However, many traditional IO models focus exclusively on quantity adjustments, leaving the role of price flexibility underexplored. 
Despite these efforts, most IO models still focus primarily on quantity adjustments, even though surveys and computational studies show that real-world firms do adjust both price and quantity in response to demand shocks \cite{bhaskarPriceQuantityAdjustment1993,kawasakiDisequilibriumDynamicsEmpirical1982,assenzaPQStrategiesMonopolistic2015,davisNominalShocksMonopolistically2011}.

In contrast to approaches that emphasize quantity adjustments, here we propose to simultaneously quantify the extent to which both price and quantity adjustments affect the output of industrial sectors in response to demand shocks. We generalize the standard IO framework by incorporating explicit mechanisms for price adaptation alongside output changes. In that model, every industrial sector that produces a homogeneous good responds to a monetary demand shock with three possible adjustments: (i) altering its production quantity, (ii) adjusting its price, or (iii) implementing a combination of both. Behavioral parameters govern the extent to which the sector responds in each dimension, determining the relative weight assigned to adjustments in price, quantity, or their combination.

\begin{figure}[htbp!]
    \centering
    \includegraphics[width=1.\textwidth]{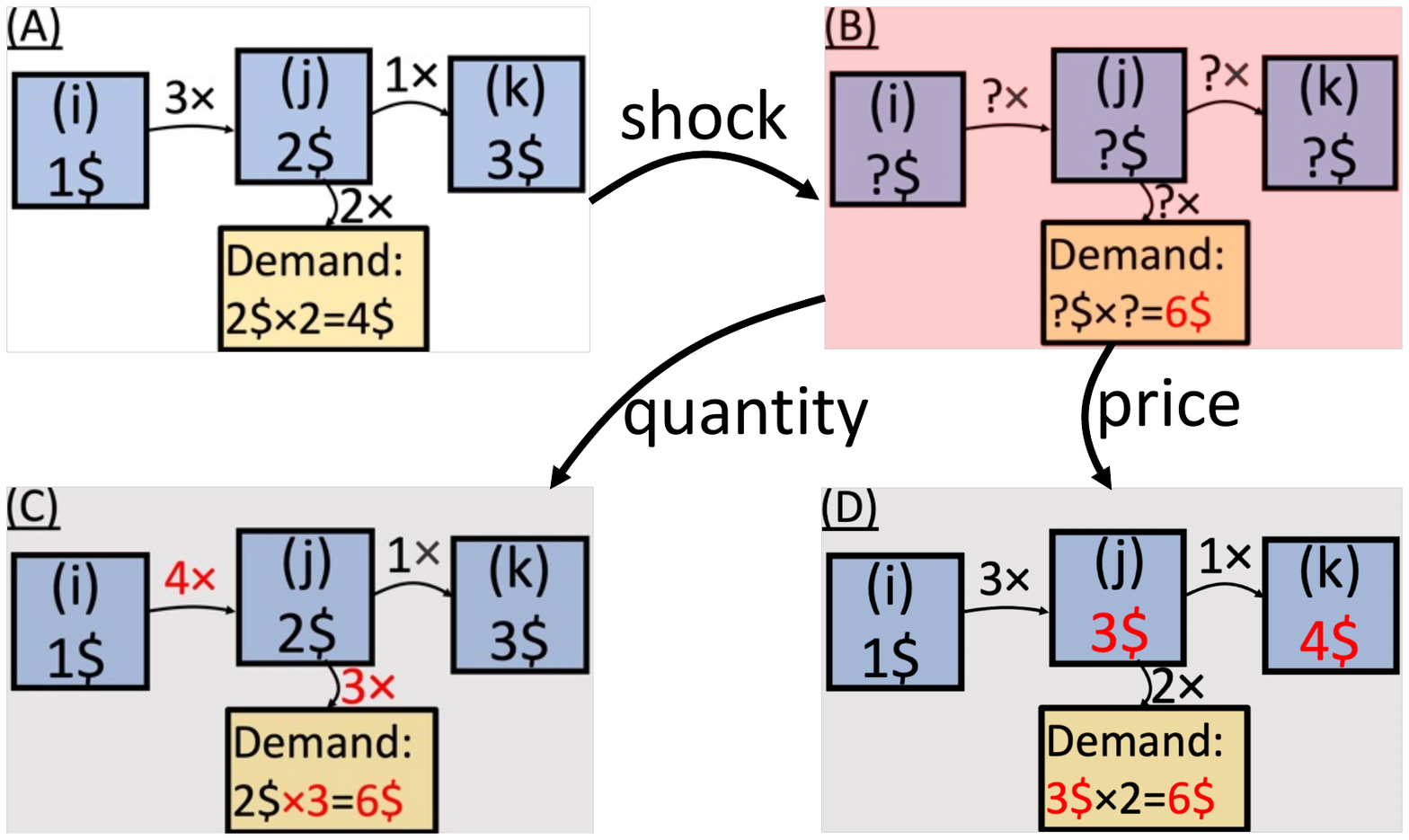}
    \caption{
    (A) Production network consisting of three ($(i)$,$(j)$,$(k)$) sectors and one end-consumer. The goods of $(i)$ are purely used by $(j)$, $(j)$ produces for $(k)$ and the demand of the end consumer.
    If $(j)$ experiences a demand shock of 2\$ (B), it can either increase its production (C) or its price (D). If the sector chooses to increase its output, this will have upstream effects, as the industry will require more goods from sector $(i)$ as inputs. 
    If sector $(j)$ decides to increase its price, it will have downstream effects as the inputs for $(k)$ increase in price.
    }
    \label{fig1}
\end{figure}

Figure~\ref{fig1} provides a stylized illustration of these dynamics using a simple network of three sectors $(i, j, k)$. Suppose that sector $j$ experiences a monetary demand shock of \$2. If $j$ responds primarily by raising output, it triggers an upstream effect: sector $i$ must supply more inputs, potentially straining its capacity. Conversely, if $j$ increases its price, it forces downstream adjustments: sector $k$ now faces more expensive inputs.

This model addresses the challenge of ensuring that the system reaches a new steady state after the shock is applied. For that we develop an iterative update process for the IO table, where we introduce behavioral adjustment parameters that govern the magnitude of price and quantity updates in each iteration (see update equations \eqref{eq:q-update}-\eqref{eq:g-update}). This approach guarantees that the system neither over- nor undershoots while it adapts to the shock. By using IO tables from the World Input–Output Database (WIOD) \cite{gaaitzenj.devriesIllustratedUserGuide}, we infer these sector-specific behavioral parameters. Specifically, we estimate the parameters for 56 sectors in 43 countries for the period 2000–2014 by comparing IO tables for consecutive years and minimizing the discrepancy between the model output and the empirically observed adjustments.

The paper is organized as follows. In Section \ref{sec: methods}, we present the formal description of the extended IO model and outline the iterative update process. Section \ref{sec:fitting} discusses the estimation strategy using WIOD data and elaborates on the calibration of sector-specific behavioral parameters. In Section \ref{sec: results}, we analyze the behavioral parameters fitted to investigate what they reveal about the inherent characteristics of sectors and countries. We explore the dimensions along which these parameters remain stable, shedding light on whether they represent intrinsic properties of specific sectors or nations. Finally, Section \ref{sec: discussion} concludes with a discussion of the findings, limitations, and avenues for future research.

\newpage
\FloatBarrier
\section{Methods} \label{sec: methods}
We now present a dynamic input-output (IO) framework that describes how industries adjust their quantity and price levels in response to demand shocks.

\subsection{Model Setup and Notation} \label{sec:setup}
To formalize the approach, consider an IO table on $n$ sectors for a specific country, $c$, in year, $y$. Let
\begin{itemize}
    \item $Z_{ij(c,y)}$ be the monetary intersectoral flow from sector $i$ to sector $j$,
    \item $f_{i(c,y)}$ be the final demand of sector $i$,
    \item $v_{i(c,y)}$ be the value-added of sector $i$, and 
    \item $x_{i(c,y)}$ be the total monetary output of sector $i$.
\end{itemize}
We assume that the typical IO consistency relationships hold,
\begin{equation}
x_k = \underbrace{\sum_{i=1}^n Z_{ik(c,y)} + v_{k(c,y)}}_{\text{Inflows}} = \underbrace{\sum_{j=1}^n Z_{kj(c,y)} + f_{k(c,y)}}_{\text{Outflows}} \quad .
\end{equation}
Next, we define \textit{price} and \textit{quantity} vectors for every sector, 
\begin{align}
    x_{i(c,y)} = q_{i(c,y)} \cdot p_{i(c,y)} = (q_{(c,y)} \odot p_{(c,y)})_i \quad ,
\end{align}
where $\odot$ denotes element-wise multiplication. Assuming that every sector produces a single homogeneous good, we define the technical coefficients, $A_{ij(c,y)}$, that describe the quantity of input from sector $i$ required to produce one unit of output in sector $j$,
\begin{align}
A_{ij(c,y)} = \frac{Z_{ij(c,y)}/p_{i(c,y)}}{q_{j(c,y)}} \quad .
\end{align}
We assume these technical coefficients are constant on the short timescale considered here. This assumption is supported by empirical studies showing that IO coefficients that reflect underlying technologies, typically change on a longer timescale~\cite{watanabeTestConstancyInputOutput1961,dietzenbacherCoefficientStabilityPredictability2006,bonComparativeStabilityAnalysis1996}.  

\subsection{Dynamical Update Equations} \label{sec:update_eqs}
To study how sectors adjust to an \textit{exogenous monetary demand shock}, we introduce an iterative process. Let $\hat{.}$ denote the variables associated with the iterative process and $\hat{.}^t$ the iteration index (not in real time), capturing the evolving states of quantity, price, and demand. 

\textbf{Initial Condition:} We start from an \textit{equilibrium state} at $t=0$ where supply equals demand.  

\textbf{Introducing a Shock:} We apply a monetary shock, $\hat{g}^0$, to one or more sectors, representing an unexpected shift in final demand. Our goal is for the system to converge to a new equilibrium in which the shock is absorbed, i.e., $\hat{g}^t \to 0$ as $t \to \infty$. During adjustment, sectors alter their \textit{quantities} $\hat{q}^t$ and \textit{prices} $\hat{p}^t$.

\textbf{Update Rules:} We propose the following set of differential equations
\begin{align}
\dot{\hat{q}}^t &= \Delta_q \big(\hat{g}^t \odot (\mathcal{P}\hat{v}^t)^{-1} \big), \label{eq:q-update}\\
\dot{\hat{v}}^t &= \Delta_p \hat{g}^t, \label{eq:v-update} \\
\dot{\hat{g}}^t &= -\mathcal{G} \dot{\hat{q}}^t \odot \mathcal{P}\hat{v}^t - \mathcal{G}\hat{q}^t \odot \mathcal{P} \dot{\hat{v}}^t, \label{eq:g-update}
\end{align}
where $\mathcal{G} = (\mathbb{I} - A)$ and $\mathcal{P} = (\mathbb{I} - A^T)^{-1} \cdot \mathrm{diag}(\hat{v}^0) \mathrm{diag}(\hat{x}^0)$.

The equations reflect iterative adjustments in the quantities, $\hat{q}^t$, value-added, $\hat{v}^t$, and the residual shock, $\hat{g}^t$.
Equation \eqref{eq:q-update} describes how sectoral quantities adjust in response to the residual demand shock, $\hat{g}^t$. The term $\big(\hat{g}^t \odot (\mathcal{P}\hat{v}^t)^{-1} \big)$ converts the residual monetary shock into physical units, using the inverse of the price vector, $(\mathcal{P}\hat{v}^t)^{-1}$. This is necessary since the left side of this equation $\hat{q}^t$ also has physical units. The parameter $\Delta_q$ determines the magnitude of these adjustments, reflecting the sectors’ behavioral inclination toward quantity changes.
Equation \eqref{eq:v-update} captures adjustments in the value-added in response to the residual monetary shock. Since $\hat{v}^t$ is already expressed in monetary terms, no conversion is needed. $\Delta_p$ is the speed and extent of this adjustment that corresponds to sectoral pricing responses.
Finally, Eq. \eqref{eq:g-update} updates the residual shock by accounting for adjustments already made in quantities and value-added. The term $-\mathcal{G} \dot{\hat{q}}^t \odot \mathcal{P}\hat{v}^t$ reflects the reduction in the shock due to changes in quantities and $-\mathcal{G}\hat{q}^t \odot \mathcal{P} \dot{\hat{v}}^t$ captures the shock reduction from changes in value-added.

Together, these terms ensure that the system moves iteratively toward a new equilibrium without overshooting.

\subsection{Fitting Procedure} \label{sec:fitting}
To estimate the behavioral parameters, $\Delta_q$, and $\Delta_p$, from empirical data, we use input-output (IO) tables from two consecutive years. The model is initialized using the IO table from the first year, and the observed change in demand between the two years is applied as a shock. Gradient descent is then used to optimize $\Delta_q$ and $\Delta_p$, ensuring that the model's resulting IO table aligns as much as possible with the observed IO table from the second year.

\subsubsection{Initialization}
Let $x_{(y1,c)}$ represent the observed total output in the first year, $y1$. We initialize:
\begin{align}
\hat{q}^0 &= x_{(y1,c)}\quad , \\
\hat{p}^0 &= (1,1,\dots,1)^T \quad , \\
\hat{A} &= \frac{Z_{ij(c,y1)}}{\hat{p}^0_{i(c,y1)} \hat{q}^0_{j(c,y1)}} = \frac{Z_{ij(c,y1)}}{x_{j(c,y1)}} \quad .
\end{align}

\textbf{Shock Definition:} The monetary shock is defined as
\begin{align}
\hat{g}^0 = f_{(c,y2)} - f_{(c,y1)} \quad .
\end{align}

\subsubsection{Residual and Optimization}
The residual, $R$, measures the deviation between observed year-2 IO values and the fully adjusted model output,
\begin{align}
R(\Delta_q, \Delta_p) = \|x_{(c,y2)} - \hat{x}^\infty\| + \|v_{(c,y2)} - \hat{v}^\infty\|.
\end{align}
We minimize $R$ to estimate $\Delta_q$ and $\Delta_p$ as 
\begin{align}
\min_{\Delta_q, \Delta_p} R(\Delta_q, \Delta_p).
\end{align}

\subsubsection{Optimization Procedure Details}
We use the \texttt{DiffEqFlux.jl} package~\cite{rackauckas2020universal} to perform the optimization. The results of the optimization process are not unique due to its stochastic nature, including sensitivity to factors such as initial values. To address this variability, we conduct 100 independent optimization runs with varying initial conditions and retain the 25 runs with the lowest residuals, $R$.  

\begin{figure}[htbp!]
    \centering
    \includegraphics[width=.5\textwidth]{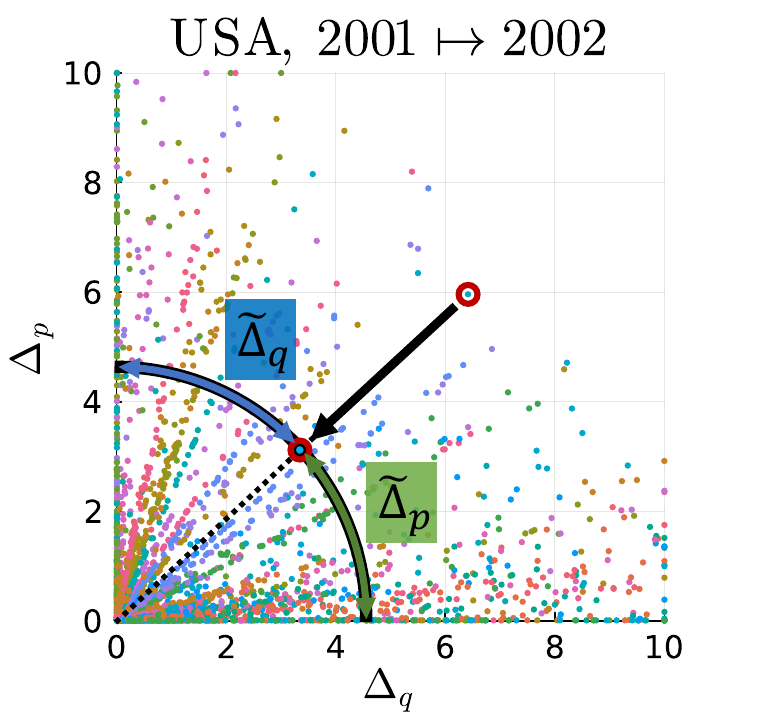}
    \caption{Scatterplot of $(\Delta_q, \Delta_p)$ pairs for 25 optimization runs in the U.S. ($2001 \mapsto 2002$).
    Each color represents a different sector, and the data points forming distinct rays originating from $(0,0)$.
    This pattern indicates that the ratio $(\Delta_q,\Delta_p)$ remains approximately constant across different optimization runs for each sector, reflecting a proportional relationship between these parameters.
    The projection onto the unit circle illustrates how the arctan transformation maps these values, effectively normalizing them into a single variable representation. The curved segments in the plot correspond to the transformed values $\tilde{\Delta}_q$, $\tilde{\Delta}_p$, showing how the original values are rescaled.
    }
    \label{fig:projection}
\end{figure}

We show a scatterplot of the parameters $(\Delta_q, \Delta_p)$ in \figref{fig:projection}. These parameters exhibit a correlation across multiple optimization runs, which arises because the fully adjusted input-output (IO) table, \(\hat{.}^\infty\) depends only on the relative magnitude of $\Delta_q$ and $\Delta_p$, not on their absolute values.  

Since $\Delta_q$ and $\Delta_p$ are interdependent, we can represent them as a single variable by projecting onto the unit circle, as  
\begin{align}
\tilde{\Delta}_p &= \frac{2}{\pi} \arctan\left(\frac{\Delta_p}{\Delta_q}\right) \quad , \\
\tilde{\Delta}_q &= \frac{2}{\pi} \arctan\left(\frac{\Delta_q}{\Delta_p}\right) \quad ,
\end{align}
with \(\tilde{\Delta}_q + \tilde{\Delta}_p = 1\). This transformation provides a unified representation of the parameters.

\newpage
\FloatBarrier
\section{Results}\label{sec: results}
We first illustrate how the behavioral parameters, $(\Delta_q,\Delta_p)$, shape shock responses in a simple four-sector toy economy (Section~\ref{sec:toy}), and then show how we estimate and interpret these parameters using real-world IO data (\ref{sec:estimates}). We then examine sector- and country-specific patterns over time and discuss their alignment with historical macroeconomic events (Section~\ref{sec:sector_country_patterns}).
Finally, we analyze the time-averaged patterns across sectors and countries (Section~\ref{sec:time averages}) and explore the correlations between them.

\subsection{A Four-Sector Toy Example} \label{sec:toy} 
\begin{figure}
    \centering
    \includegraphics{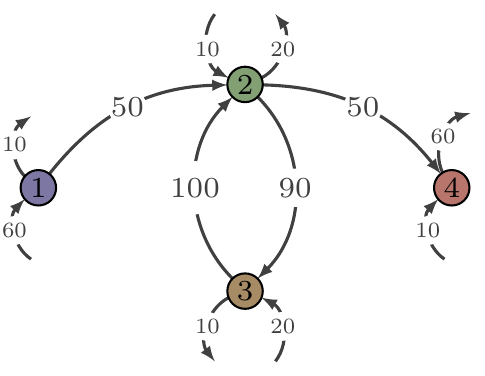}
    \caption{Network representation of a four-sector toy economy. Nodes denote sectors, arrows indicate monetary flows. Value added and demand are represented as external inflows/outflows. The visualization highlights the balanced in-  and outflow for each sector.}
    \label{fig:toy_graph}
\end{figure}

\begin{figure}[htbp!]
    \centering
    \includegraphics[width = 1\linewidth]{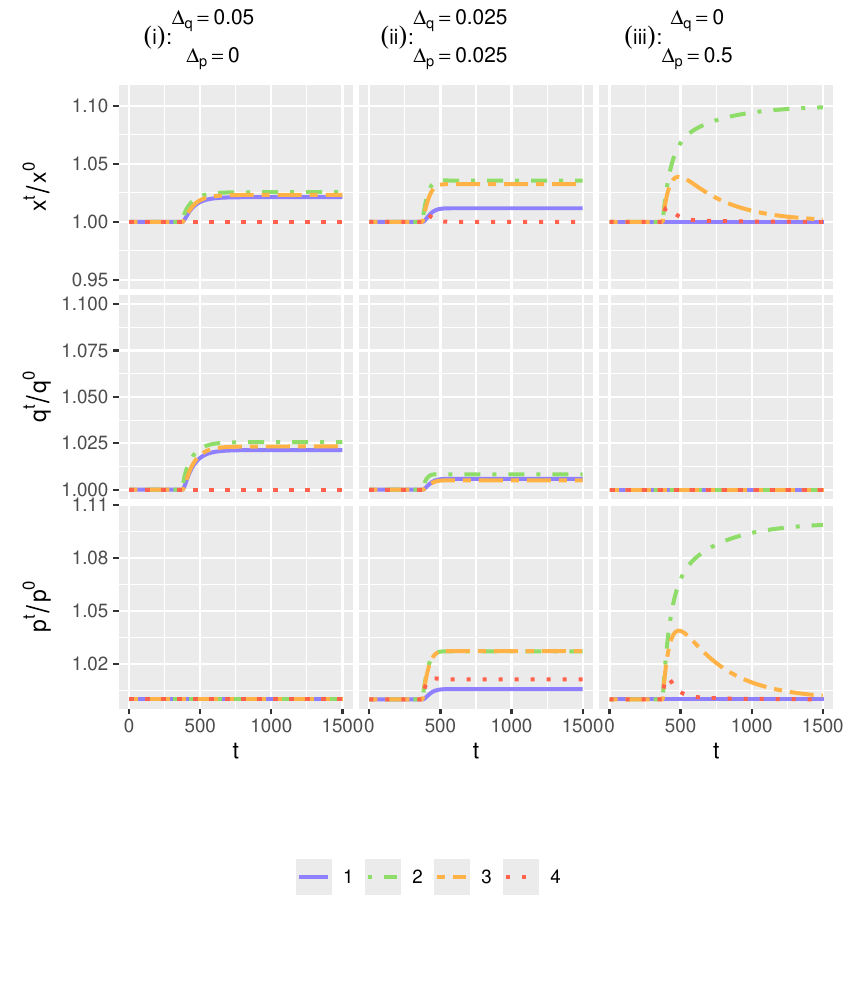}
    \caption{Trajectories of output, ${x_t}/{x_0}$, quantity, ${q_t}/{q_0}$, and price ${p_t}/{p_0}$ of the model. A demand shock of 10\% in sector 2 is introduced at $t=250$. The different colors and lines represent the four sectors. Each column represents different values of the behavioral parameters, $\Delta_q$ and $\Delta_v$.}
    \label{fig:toy trajectories}
\end{figure}

To illustrate the model we present a simple four-sector toy economy, represented by the network shown in \figref{fig:toy_graph}. Arrows represent intersectoral monetary flows. Value-added and final demand for each sector (nodes) are represented by arrows originating from or terminating elsewhere (not at the nodes), reflecting external sources or sinks. The network maintains a balance between total inflow and outflow for each sector at each point in the model dynamics.
\figref{fig:toy trajectories} illustrates the dynamic response of a toy economy to a 10\% demand shock in sector 2. Initially at equilibrium, the shock is introduced at t=250. The figure shows the trajectories of the output (${x_t}/{x_0}$), the quantity (${q_t}/{q_0}$), and the price (${p_t}/{p_0}$), of the model, relative to the values at the initial equilibrium state.

The columns in \figref{fig:toy trajectories} show the trajectories for different sets of behavioral parameters, $(\Delta_q,\Delta_p)$.
The first column, $(i)$, shows exclusive quantity adjustment, the second, $(ii)$, a mixed price and quantity adjustment, and the third, $(iii)$, displays exclusive price adjustments. 
For the case of exclusive quantity adjustments in sector $i$, these quantity adjustments propagate upstream while the downstream quantities and all prices remain constant.
For exclusive price adjustments, all quantities and upstream prices remain constant, while price changes propagate downstream.
In the mixed case, consequently, prices and quantities might change in all sectors.
The toy model demonstrates that these processes can result in different sectoral responses to the shock in terms of output.

\subsection{Estimated adjustment parameters across Sectors} \label{sec:estimates}
\begin{figure}[htbp!]
    \centering
    \includegraphics[width = 1.\textwidth]{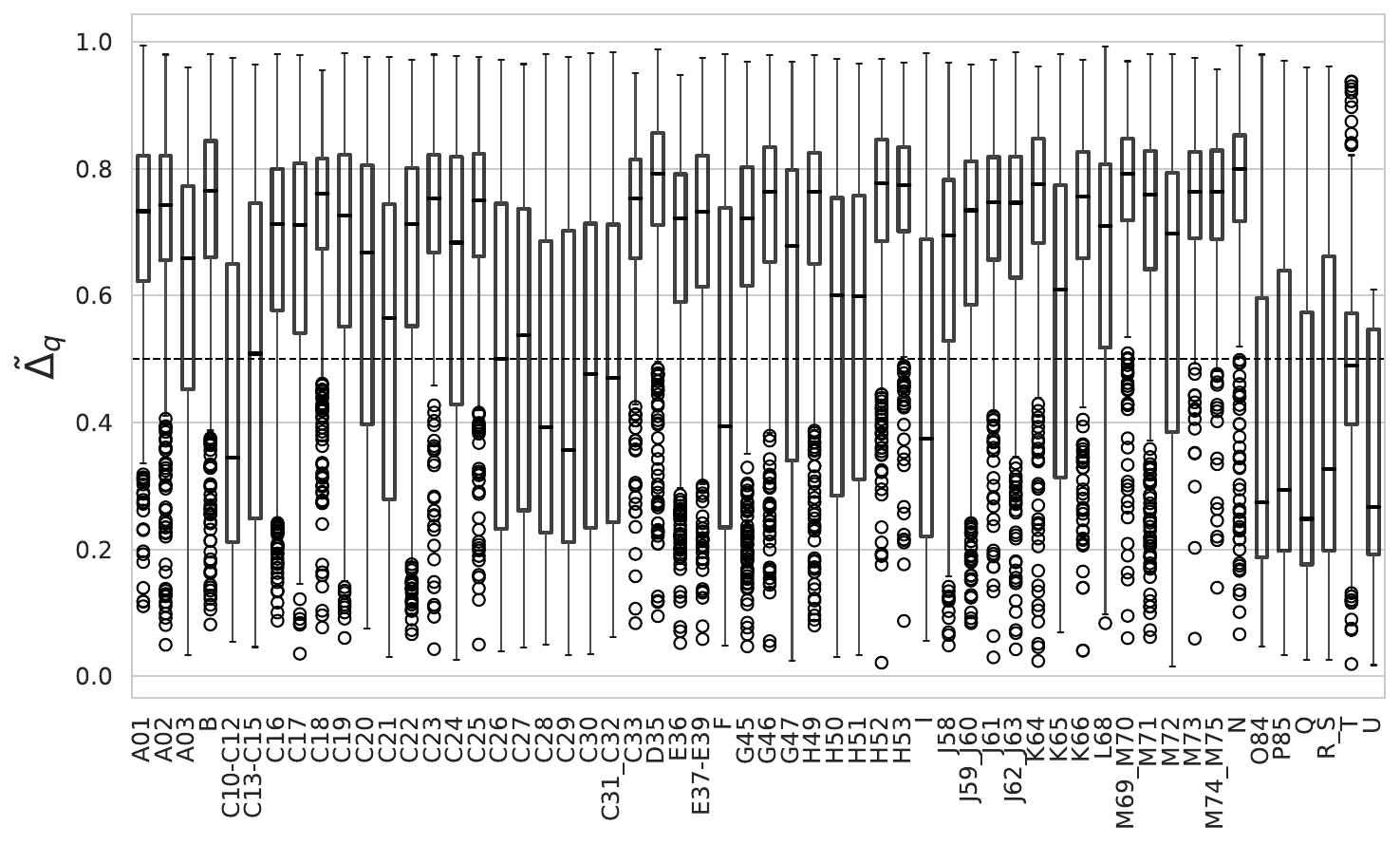}
    \caption{
    Boxplot of the fitted $\tilde{\Delta}_q$s, stratified by sector. The parameters show a large variation and, in general, can be found between 0 and 1.
    For most sectors, the median, and for most, even the lower limit of the interquartile range lies above $\tilde{\Delta}_q > 0.5$, indicating that most sectors tend to adjust quantity.
    }
    \label{fig:boxplot sectors}
\end{figure}

We now apply the model to consecutive-year IO tables from the World Input–Output Database (WIOD)~\cite{gaaitzenj.devriesIllustratedUserGuide} and assume that the observed annual changes result primarily from external demand shocks. By minimizing the difference between model outcomes and empirical IO data, we obtain fitted values for $\tilde{\Delta}_q$, for every sector.
\figref{fig:boxplot sectors} shows the distribution of fitted $\tilde{\Delta}_q$ values for every sector.
Note the considerable variation, $\tilde{\Delta}_q$ can be between 0 and 1 for most sectors.
However, it was observed that the majority of sectors had a median and lower interquartile range limit greater than $\tilde{\Delta}_q > 0.5$, suggesting that, overall, sectors tend to favor quantity over price adjustments.
There are, however, sectors that don't follow this general trend. For instance, for sectors C27 (Manufacture of electrical equipment) and C28 (Manufacture of machinery and equipment ) most $\tilde{\Delta}_q$ values are below $0.5$, indicating a tendency towards price adjustments.

\subsection{Sectoral and Country-Specific Patterns Over Time} \label{sec:sector_country_patterns}
\begin{figure}[htbp!]
    \centering
    \includegraphics[width = .9\textwidth]{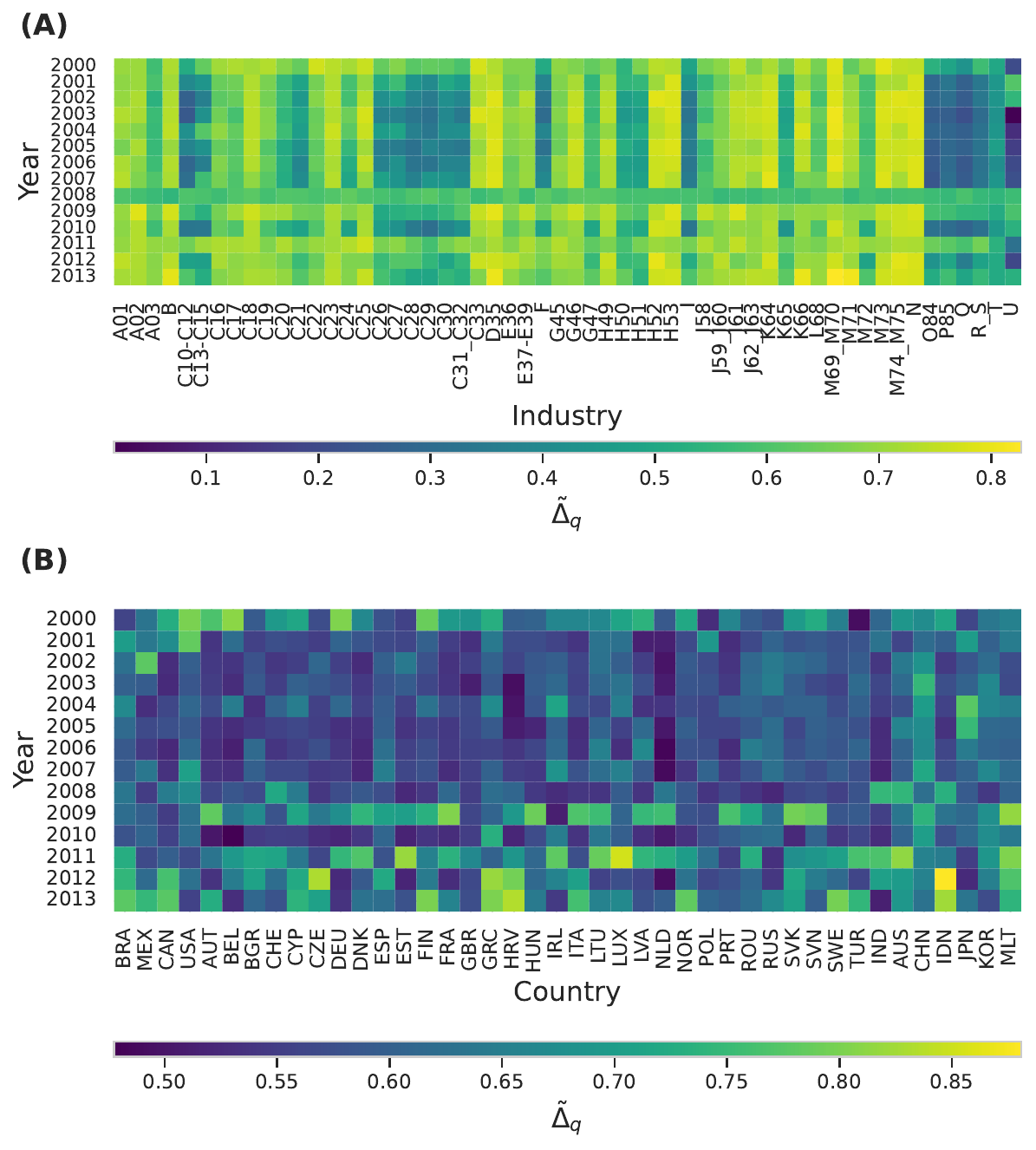}
    \caption{
     Evolution of average $\tilde{\Delta}_q$ for various sectors (A) and countries (B) between 2000 and 2013.  In both cases, there is a broadly consistent pattern in the \emph{relative} adjustment preferences across sectors and countries year-over-year, with notable exceptions during the economic shock years of 2000, 2008, 2009, and 2011. The structure of $\tilde{\Delta}_q$ values for specific sectors appears more consistent across different sectors than the structure observed across different sectors over time. This suggests that $\tilde{\Delta}_q$ primarly reflects sector-specific properties rather than country-specific ones.}
    \label{fig:heatmap year x}
\end{figure}

To examine how $\tilde{\Delta}_q$ evolves over time and between regions and sectors, we aggregate sector-level estimates annually and present the results in Figure~\ref{fig:heatmap year x}. Although the overall structure of $\tilde{\Delta}_q$ remains consistent over the years, the sectoral averages (panel (B)) reveal notable outliers during specific years: 2000, 2008, 2009, and 2011. These years align with significant economic disruptions, such as the dot-com bubble~\cite{dot-com}, the financial crisis~\cite{2007-crisis}, and the European debt crisis~\cite{european-crisis}.

The country-level averages (panel (A) in Figure~\ref{fig:heatmap year x}) provide additional insights: For countries in the Americas, such as Brazil, Mexico, Canada, and the United States, $\tilde{\Delta}_q$ values had largely returned to normal by 2009. In contrast, many European countries experienced a prolonged recovery, highlighting the lingering impact of the European debt crisis.

These patterns suggest that the behavioral parameters, $\tilde{\Delta}_q$, are an inherent property of the economy, reflecting its structural and regional dynamical characteristics.

\subsection{Time Averages}\label{sec:time averages}
\begin{figure}[htbp!]
    \centering
    \includegraphics[width = 1.\textwidth]{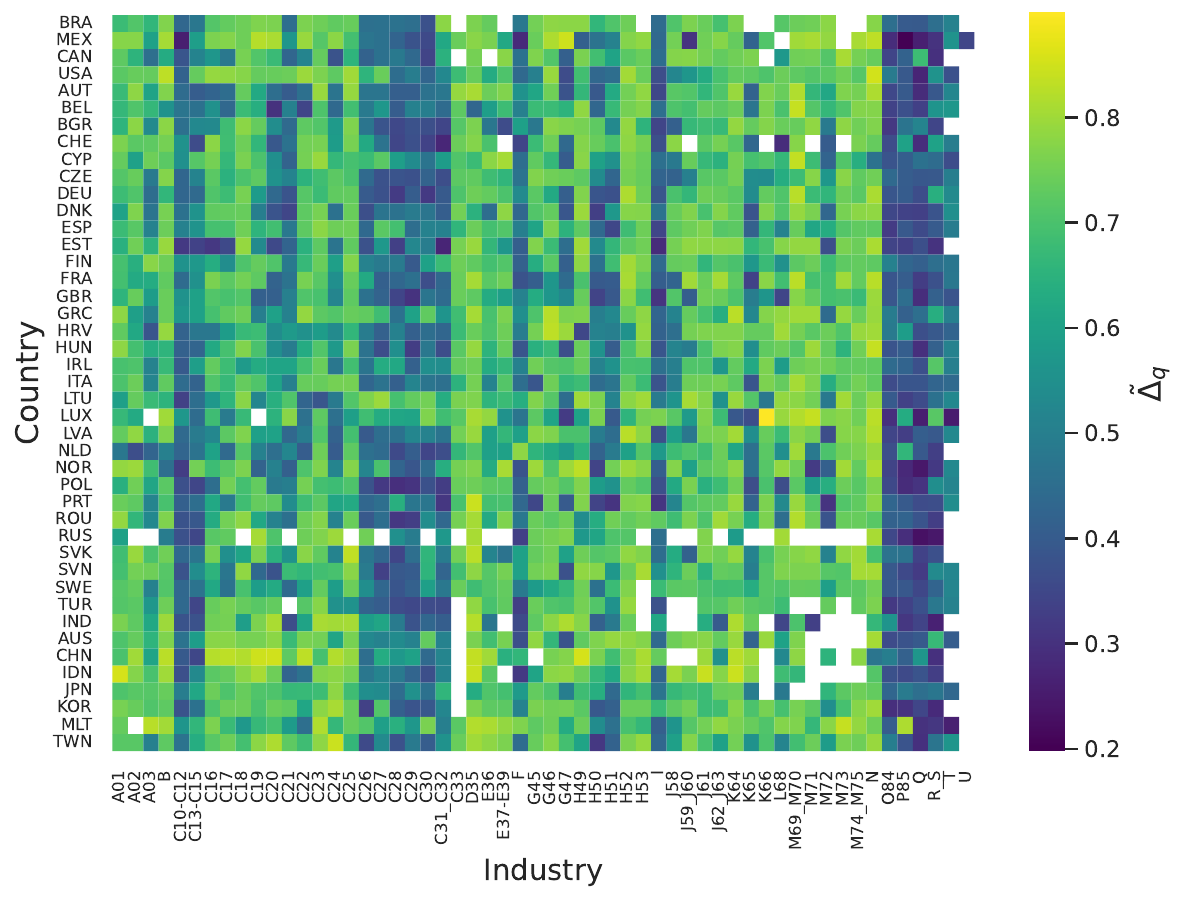}
    \caption{
        $\tilde{\Delta}_q$, stratified by sectors and countries. Countries are grouped by continent (Latin America \& Caribbean, North America, Europe \& Central Asia, South Asia, East Asia \& Pacific, Middle East \& North Africa). 
        Countries in the same regions exhibit similar adjustment behavior; for instance, European countries share remarkable similarities, as do Asian countries.
    }
    \label{fig: heatmap industry x country}
\end{figure}

Figure~\ref{fig: heatmap industry x country} shows the time-averaged $\tilde{\Delta}_q$ for various industries (A) and countries (B). A clear pattern emerges: European countries, with the exception of Switzerland and Russia, exhibit a similar structure in their sector-specific $\tilde{\Delta}_q$ values. Asian countries also share a comparable structure within their region. The differences between European and Asian countries are evident.  
European countries tend to show higher values $\tilde{\Delta}_q$ in specific manufacturing sectors, including C16 (Wood Products), C19 (Coke and Refined Petroleum), C20 (Basic Chemicals, Fertilizers, and Nitrogen), and C21 (Basic Pharmaceuticals). This regional divergence is also visible in Figure~\ref{fig:heatmap correlation countries(industries)} in the Appendix, which shows the correlations of country-specific $\tilde{\Delta}_q$ values across sectors.
The similarity of manufacturing industries is seen in the higher correlation in Figure~\ref{fig:heatmap correlation industries(countries)} in the Appendix, which shows correlations of sector-specific $\tilde{\Delta}_q$ values across countries.

\FloatBarrier

\section{Discussion}\label{sec: discussion}
We introduce a new approach to the analysis of economic shocks based on three interconnected update equations that describe how quantities and prices adjust simultaneously in response to shocks. These changes lead to a new equilibrium. The presented framework extends the foundational Leontief quantity and price models in input-output analysis in a natural way. The Leontief quantity model, which assumes fixed prices, examines how changes in final demand affect sector outputs, while the price model, which assumes fixed quantities, explores how changes in value-added components, such as wages or profits, influence prices~\cite{dietzenbacher1997vindication,millerInputOutputAnalysisFoundations2009}.

In the presented model, quantities and prices adjust simultaneously, reflecting their dynamical interaction. The quantity update equation captures how quantities respond to demand shocks while incorporating the influence of price changes. Similarly, the price update equation demonstrates how prices adjust to shocks, accounting for changes in quantities. Finally, the shock update equation integrates these adjustments to ensure that the system reaches an equilibrium without over- or undershooting.
We extend previous models that omitted prices entirely~\cite{contrerasPropagationEconomicShocks2014,klimekQuantifyingEconomicResilience2019a,pichler21,haimesInoperabilityInputOutputModel2005} or modeled them through supply-demand relationships~\cite{hallegatteAdaptiveRegionalInputOutput2008} or profit maximization~\cite{mandelEconomicCostCOVID2020}. 
Here, price formation is directly embedded into the IO structure. 
Rather than direct market clearing, prices adjust implicitly based on how changes in sectoral value-added, driven by demand shocks, ripple through the input-output network. This value-added driven mechanism determines the price level required to maintain consistency within the IO accounts, given the quantity adjustments and the fixed technical coefficients. 
This particular realization offers tractability but simplifies real-world pricing; the general framework, however, could potentially incorporate more complex pricing rules via modified update equations.

This endogenous approach enables us to estimate and fit sector-specific behavioral parameters that characterize whether sectors tend to prefer quantity over price adjustments or the other way round, allowing us to find strong and consistent regional and sector-specific behavioral patterns.

This analysis reveals how different regions respond differently to demand shocks and thus realize economic resilience and recovery dynamics in distinct ways. 
Using the World Input-Output Database, we found characteristic regional adjustment patterns: European countries generally exhibit a similar structure in their sector-specific preferences for quantity versus price adjustments $(\tilde{\Delta}_q)$, a pattern also observed within Asian countries, though the typical profiles differ between these two regions. Notably, European economies often show a stronger tendency towards quantity adjustments $(\tilde{\Delta}_q>0.5)$ in specific manufacturing sectors like Wood Products, Coke/Refined Petroleum, Chemicals, and Pharmaceuticals compared to other regions. 
These findings suggest that the preference for price or quantity adjustments holds consistent characteristics both within certain geographic regions and across specific types of industries globally.

Comparing these empirical findings with previous research, some studies have explored combined price and quantity strategies using computational market models, though often without direct calibration to macro-level data~\cite{assenzaPQStrategiesMonopolistic2015,davisNominalShocksMonopolistically2011, gatti2011macroeconomics}. 
Our observation that sectors generally favor quantity adjustments aligns with earlier survey-based evidence showing firms are more likely to adjust output instead of prices in response to demand shocks~\cite{kawasakiDisequilibriumDynamicsEmpirical1982}. 
This consistency is particularly notable given our finding that $\tilde{\Delta}_q > 0.5$ holds true for most sectors across the dataset, even while exhibiting the distinct regional variations highlighted above.

Fitting the behavioral parameters required assuming a separation of time scales, namely that the model reaches a new steady state reflecting the end-of-year conditions annually. This approach, while necessary for leveraging annual IO data, means the model lacks explicit timescales and cannot predict the economy's state at a specific future point in time. This limits its predictive power in comparison to other dynamic models~\cite{pichler21,klimekQuantifyingEconomicResilience2019a}. 
We also use a highly stylized input-output model, which assumes that all firms in a sector can be represented by one representative firm, which is certainly unrealistic ~\cite{diemEstimatingLossEconomic2023}.
Although here we applied the model only  to country-wide, industry-level IO tables, it could be potentially used with much more granular datasets, such as firm-level supply networks.

In conclusion, this work introduces a novel extension to dynamic input-output models by incorporating simultaneous price and quantity adjustments governed by empirically derived behavioral parameters. By examining their correlation over time, sectors, and countries, we find that these parameters reveal distinct sector- and country-specific traits. This framework moves beyond traditional fixed-price or fixed-quantity assumptions, offering a more nuanced approach to analyzing how economies absorb and recover from shocks.

\subsection*{Acknowledgements}
On behalf of the Complexity Science Hub Vienna (CSH), we acknowledge funding from the Österreichische Nationalbank (OeNB) in the framework of the Austrian Economic Research Promotion Programme, and from the Austrian Research Promotion Agency (FFG) through the ESSENCSE project (No. 873927). On behalf of the Supply Chain Intelligence Institute Austria (ASCII), we acknowledge financial support from the Austrian Federal Ministry for Economy, Energy and Tourism (BMWET) and the Federal State of Upper Austria.

\FloatBarrier
\newpage
\bibliography{lit.bib, lit_manual.bib}

\FloatBarrier
\newpage
\section*{Appendix}
\begin{figure}[htbp!]
    \centering
    \includegraphics[width = .9\textwidth]{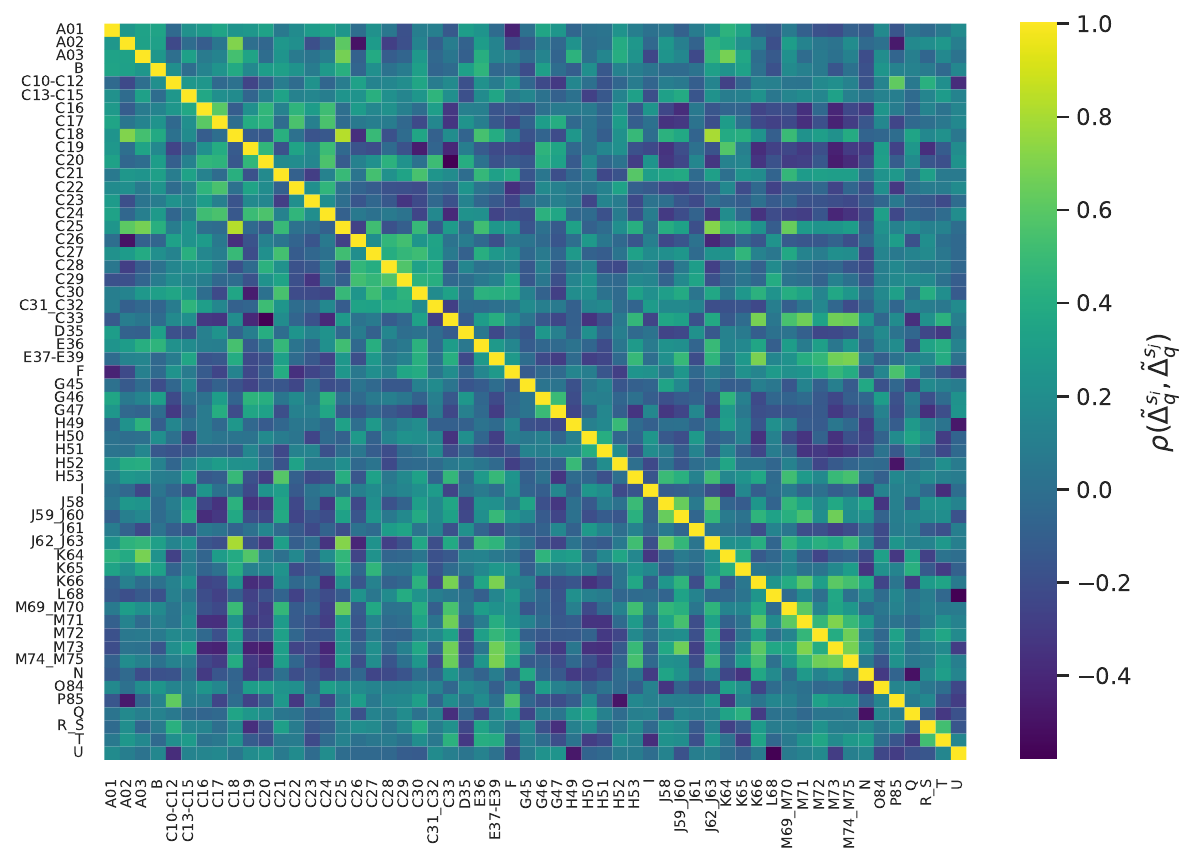}
    \caption{Correlation of $\tilde{\Delta}_q$ of the different sectors over the countries. This reveals a clustering of the Manufacturing sectors C25-C30
    }
    \label{fig:heatmap correlation industries(countries)}
\end{figure}

\begin{figure}[htbp!]
    \centering
    \includegraphics[width = 1.\textwidth]{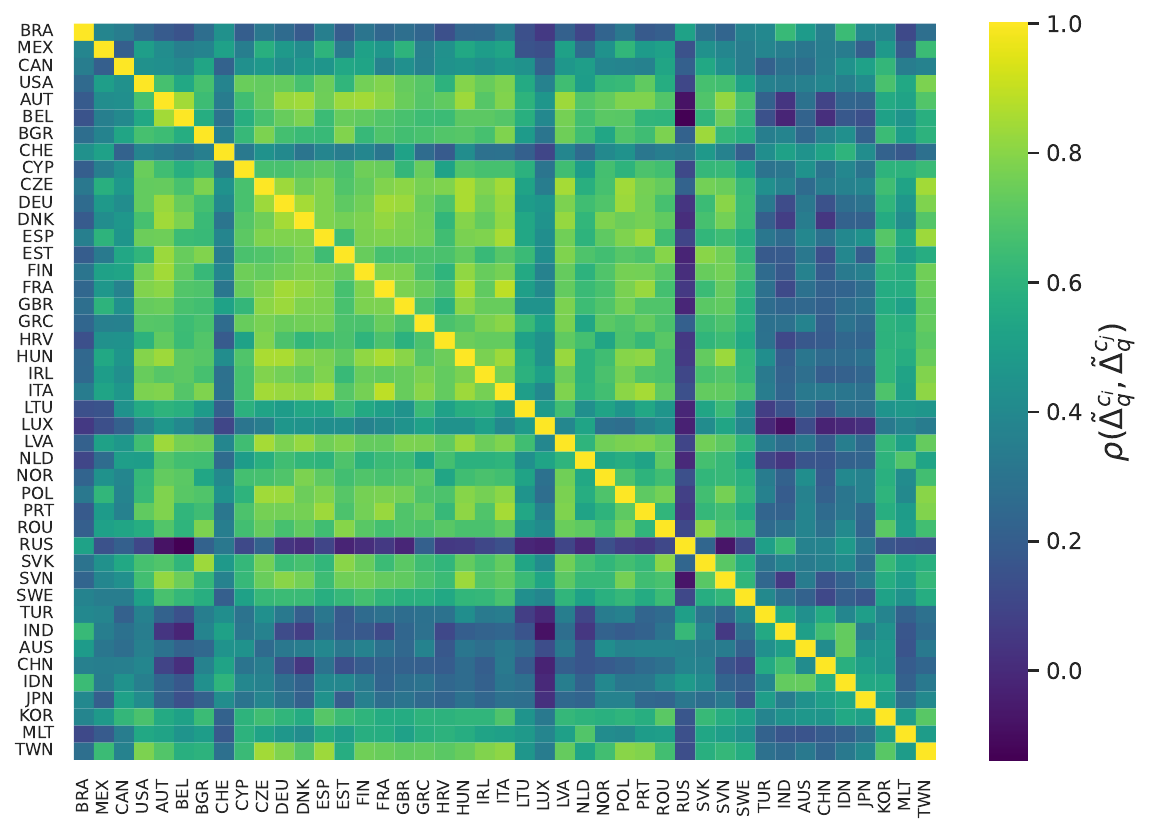}
    \caption{Correlation of $\tilde{\Delta}_q$ of the different countries over the sectors.
    The European countries are very clustered (exception of Switzerland and Russia) and the Asian-Oceanian countries as well (including Turkey). }
    \label{fig:heatmap correlation countries(industries)}
\end{figure}

\end{document}